\documentclass[aps,twocolumn]{revtex4}

\usepackage{amsmath}
\usepackage{amsfonts}
\usepackage{amssymb}
\usepackage{mathrsfs}
\usepackage{graphicx}
\usepackage{subfigure}

\newcommand{\bs}[1]{\boldsymbol{#1}}

\begin{document}
 \title{Coherence and correlation functions of quasi-2D dipolar superfluids at zero temperature}

\author{Andrew G. Sykes}
\author{Christopher Ticknor}
%\thanks{}
\affiliation{Theoretical Division and Center for Nonlinear Studies, Los Alamos
National Laboratory, Los Alamos, NM 87545, USA}

\begin{abstract}
 We use the Bogoliubov theory of Bose-Einstein condensation to study the properties of dipolar 
particles (atoms or molecules) confined in a uniform two-dimensional geometry at zero temperature. We find 
equilibrium solutions to the dipolar Gross-Pitaevskii equation and the Bogoliubov-de Gennes equations. 
Using these solutions we study the effects of quantum fluctuations in the system, particularly 
focussing on the instability point, where the roton feature in the excitation spectrum touches zero. 
Specifically, we look at the behaviour of the noncondensate density, the phase fluctuations, and 
the density fluctuations in the system. Near the instability, the density-density correlation function shows a 
particularly intriguing oscillatory behaviour. Higher order correlation functions display a distinct 
hexagonal lattice pattern formation, demonstrating how an observation of broken symmetry can 
emerge from a translationally symmetric quantum state.
\end{abstract}

\date\today

\maketitle

\section{Introduction}

The field of ultra-cold atoms is often heralded as an ideal system to solve outstanding 
problems in many-body physics, owing to the innate tunability of certain crucial 
parameters within the Hamiltonian. The extremely low densities and temperatures associated with 
such systems typically ensure that interactions between atoms are completely dominated by 
$s$-wave collisions. Techniques for varying the strength of such interactions, 
via Feshbach resonances, are long-standing~\cite{KetterleFeshbachDemonstration, FeshbachReview}. 
Exciting recent developments involve 
the introduction of an entirely new class of interaction, the dipole-dipole interaction, 
by cooling atoms with a large magnetic dipole moment, such as chromium or dysprosium, to 
temperatures below quantum degeneracy~\cite{ChromiumCondensate, ChromiumDipolarEffects1, ChromiumDipolarEffects2,
DysprosiumCondensate, ErbiumCondensate, DipolarBosonsReview}. Even 
more ambitious, are the attempts to cool heteronuclear molecules, which typically involve 
considerably larger dipole moments~\cite{DebbieJinKRbMolecules, RbCsMolecules}. Such systems typically have permanent electric 
dipole moments. The energy associated with the dipole interactions of heteronuclear molecules 
can be extremely large, greatly exceeding the thermal kinetic energy, or any other type of interactions in the 
system. These exciting experiments suggest that the observation of strongly correlated regimes in 
degenerate dipolar gases are close at hand. 
Another recent experiment, also aimed at exploring many-body physics beyond $s$-wave interactions, 
demonstrated the feasibility of introducing angular dependency into the interactions 
by dressing the atomic states with light~\cite{SpielmanSyntheticPartialWaves}, effectively 
generating screened interactions 
analagous to electron fluids in metals. These experiments and others of a similar vein, 
motivate the investigation into the novel physics which potentially could be observed in 
systems with long-range and angularly dependent interactions. 

In this article, we investigate a quasi-2D Bose gas at zero temperature using the 
Bogoliubov theory of Bose-Einstein condensation. While the formalism is general to interactions of any 
long range or anisotropic form, we focus on the specific case where the dipoles 
are aligned perpendicular to the plane of kinetic freedom~\cite{FischerTwoDimDipoleStability}. 
We note the formation of a roton in the excitation spectrum, due to a competition between the trapping potential 
and the attractive interaction in that direction. This roton introduces an instability in the 
system (the energy of the roton mode becomes equal to the energy of the ground state) 
at a critical value of the parameter; $a_{\rm dd}n l_z$, where 
$a_{\rm dd}$ is the dipole length scale (as defined in Ref.~\cite{DipolarBosonsReview}), 
$n$ is the projected 2D density (that is, the 3D 
particle distribution projected down onto two dimensions), and $l_z$ is the width of the density 
profile in the trapped direction. 
We pay particular interest to the implications this roton has on the 
quantum fluctuations in the system. We do so quantitatively by calculating the 
noncondensate fraction, the momentum distribution, the phase fluctuations, 
and the density fluctuations, as one approaches the critical point. 
Contrary to what might be expected from a theory which is essentially little more than an 
extension to mean field theory, we find intriguing predictions for the behaviour of 
the density fluctuations in the system near the critical point. The results suggest the 
emergence of a locally ordered state, in which density maxima lie on a hexagonal Bravais lattice. 
Although these predictions are qualitative, they give us clues regarding important concepts such as: 
How does the broken translational symmetry in a solid emerge from a quantum fluid? 
Such questions regarding symmetry breaking and pattern formation are of interest in a wide 
range of disciplines, from cosmology to biology.

A significant amount of recent theoretical focus has emerged for dipole-dipole 
interactions in systems of either bosons or fermions. In Ref.~\cite{BissettNormalStability}, the Bose gas in its 
normal state is considered, and a semiclassical theory of the system above the 
transition temperature was used to develop a qualitative picture of the stability boundaries 
for various trapping geometries. Similar work was done in Refs.~\cite{ChineseNormalFermiGas,BaillieNormalFermiGas} for 
fermionic gases. At temperatures near quantum degeneracy, the question of Cooper-pair formation, and superfluidity, 
in dipolar fermionic systems was considered in Ref.~\cite{DipolarCooperPairs,DipolarFermionicSuperfluid}. 
Inspired by traditional condensed matter achievements, it was theoretically shown how one may obtain fractional quantum 
Hall states, and realisations of the Laughlin wavefunctions, in rapidly rotating dipolar Fermi 
gases~\cite{DipolarFermionsQuantumHall}. Progress on low temperature (condensed) Bose systems has 
also been forth-coming, with a great deal of focus on numerical methods for efficiently solving 
nonlocal, nonlinear partial differential equations~\cite{BlakieDipolarGPENumerics,OtherEfficientDipolarNumerics}. 
Such equations appear in the theory of the zero-temperature analysis of the condensate 
wavefunction~\cite{DipolarBosonsReview}. Numerical solutions to the Bogoliubov-de Gennes equations in a trapped 
3D gas were also studied~\cite{RonenDipolarCylindricalTrapBogoliubovModes}. Such solutions yield a direct understanding 
of the excitation spectra and quantum depletion of such condensates. A point of interest in this work was the 
discovery of a roton-maxon in the excitation spectrum~\cite{SantosRoton,WilsonManifestationsOfDipolarRotonMode}, 
as found in the neutron-scattering experiments with 
superfluid helium~\cite{PalevskyNeutronScatteringHelium}. 
According to Landau's theory of superfluidity, this roton mode has a profound effect on the superfluid critical 
velocity of the system. This effect was studied in detail in Ref.~\cite{WilsonDipolarCriticalVelocity} using 
numerical solutions to the dipolar-Gross-Pitaevskii equation, with an object moving through the fluid. The 
results provide a powerful numerical validation of Landau's theory. The anisotropy of this critical velocity was studied 
in Ref.~\cite{TicknorAnisotropicSuperfluidity}. Topological excitations of the condensate mode have also been 
studied~\cite{YiPuVorticesDipolarBosons,WilsonVortexDipolarBosons,
AnisotropicSolitonDipolarCondensates,TwoDimBrightSolitionsDipolarCondensates}. 
This yields the structure and stability of vortices and solitons in such sytems. 

The problem of strongly correlated dipolar systems is a challenging one~\cite{AstrakharchikMeanFieldLimitations, 
JainMonteCarloDipoles}. 
It has long been understood, within the theory of classical 2D molecular dynamics, 
that interactions of the form $U(r)= D/r^3$ exhibit thermal melting transitions 
when the parameter $\Gamma=D/(k_{\rm B}Tr_0^3)$ (where $r_0$ is the average interparticle separation, 
$T$ is temperature, and $k_{\rm B}$ is Boltzmann's constant) drops below a critical value, 
approximately given by $\Gamma_c\simeq 62\pm3$~\cite{KaliaVashishta}. 
More recently, it was pointed out that, at low temperatures, quantum fluctuations 
can also melt the system~\cite{ZollerStronglyCorrelatedDipoles, AstrakharchikPhaseTransition}. When the ratio, $r=E_{\rm I}/E_{\rm K}$, 
of average interaction energy; $E_{\rm I}=D/r_0^3$, to 
zero-point kinetic energy; $E_{\rm K}=\hbar^2/(mr_0^2)$ (where $m$ is the mass of the particle), 
drops below the critical value; $r_c\simeq18\pm4$. The results of both the molecular dynamics simulations at 
finite temperature, and the path-integral Monte Carlo method at zero temperature, indicate that the 
transition is first order, which suggests the possibility of a {\it mixed phase} regime. 

Both the classical and quantum phase transitions occur when the average interaction energy 
is large compared with the average kinetic energy. For this reason, theoretical attempts to understand the 
phase-transition rely on computationally intensive algorithms such as diffusion-Monte-Carlo, or 
path-integral-Monte-Carlo. While these calculations provide an exact treatment, we demonstrate 
that the considerably simpler Bogoliubov theory shows definite presursors to the transition, where 
a lattice structure begins to emerge in the many-body density correlations, inspite of a uniform 
density remaining in the ensemble average.

Our paper is organised as follows: In Sec.~\ref{SECformalism} we introduce the formalism and notation which 
we use throughout the manuscript. The formalism can be described as an extension of Bogoliubov's 
theory of Bose-Einstein condensation to account for long range and possibly anisotropic interactions. 
In Sec.~\ref{SECexcitation} we analyse the excitation spectrum of the system, paying particular attention to the formation 
of the roton. We identify the single parameter which quantifies the strength of the roton in the system 
and the critical point at which the roton touches zero, and the system becomes unstable. 
In Sec.~\ref{SECnoncondensate} we study the noncondensate fraction as a function of the roton strength. 
We find this noncondensate fraction diverges as one approaches the critical point. In 
Sec.~\ref{SECmomentum} we study the momentum distribution of the system, and find a rapid increase in the 
population of particles occupying the roton modes as one approaches the critical point. 
In Sec.~\ref{SECphase} we study the phase coherence as a function of roton strength. In Sec.~\ref{SECdensity} 
we study the density fluctuations as a function of roton strength. Finally, we conclude in Sec.~\ref{SECconclusions}.

\section{Formalism}\label{SECformalism}
We start with the three-dimensional Hamiltonian for a bosonic field interacting via dipole-dipole 
interactions
\begin{equation}\label{EQhamiltonian}
 \hat{H}=\int_{\mathbb{R}^3}\!\! d^{3}\mathbf{r}\;\hat{\Psi}^\dagger(\mathbf{r})\left[
\frac{-\hbar^2}{2m}\nabla_{\mathbf{r}}^2+\frac{m}{2}\omega_z^2 z^2+V_{\rm int}(\mathbf{r})
\right]
\hat{\Psi}(\mathbf{r})
\end{equation}
where $V_{\rm int}(\mathbf{r})=
\frac{1}{2}\int
d^3\mathbf{r}'\hat{\Psi}^\dagger(\mathbf{r}') U(\mathbf{r}-\mathbf{r}')
\hat{\Psi}(\mathbf{r}')$ describes the two-body interactions. The field operator 
$\hat{\Psi}$ satisfies the usual, equal-time, bosonic commutation relations;
$\left[\hat{\Psi}(\mathbf{r}),\hat{\Psi}^\dagger(\mathbf{r}')\right]=\delta(\mathbf{r}-\mathbf{r}')$, 
and $\left[\hat{\Psi}(\mathbf{r}),\hat{\Psi}(\mathbf{r}')\right]=
\left[\hat{\Psi}^\dagger(\mathbf{r}),\hat{\Psi}^\dagger(\mathbf{r}')\right]=0$. 
The two-body interaction 
potential is given by
\begin{equation}
U(\mathbf{r})=\frac{D}{4\pi}\left[\frac{1-\frac{3z^2}{r^2}}{r^3}\right],
\end{equation}
where $D$ is the dipolar coupling constant, and we have assumed that some 
external field aligns all dipoles along the $z$ axis. 
The equation of motion for the Bose field operator is found from the Heisenberg equation 
of motion,
\begin{equation}\label{EQHeisenberg3D}
 i\hbar\partial_t\hat{\Psi}=\left[\frac{-\hbar^2}{2m}\nabla_{\mathbf{r}}^2+
\frac{m}{2}\omega_z^2 z^2+V_{\rm int}(\mathbf{r})\right]\hat{\Psi}.
\end{equation}
This equation is greatly simplified if we assume the field operator can be factorized into 
axial (along the $z$ axis), and radial (in the $x$-$y$ plane) components; 
$\hat{\Psi}(\mathbf{r})=\hat{\psi}(\bs{\rho})\phi_0(z)$, where $\bs{\rho}=(x,y)$. 
The form of $\phi_0$, in general will depend on 
the strength of the confining potential $\omega_z$. For instance, if the confinement is very weak, such 
that $\hbar\omega_z\ll\mu$ (where $\mu$ is the chemical potential of the system) then $\phi_0$ is well 
approximated by an inverted parabola~\cite{ODellDipolarHydrodynamics, EberleinThomasFermi} 
of width $l_{\rm TF}=\sqrt{2\mu/m\omega_z^2}$ (the Thomas-Fermi profile). 
In the opposite limit, 
where $\hbar\omega_z\gg\mu$ the system is purely 2D, and $\phi_0$ is well approximated by the 
ground state harmonic oscillator wavefunction, with width $l_{\rm HO}=\sqrt{\hbar/m\omega_z}$. 
In this manuscript, for simplicity, we assume $\phi_0$ is given by a 
Gaussian profile, of width $l_z$. In general this value of $l_z$ can be determined 
by minimising the energy, using the ansatz $\phi_0(z)=(l_z\sqrt{\pi})^{-1/2}e^{-z^2/2l_z^2}$, and depending 
on the ratio $\hbar\omega_z/\mu$, one will find $l_{\rm HO}\leq l_z\leq l_{\rm TF}$. We expect this 
theory to be quantitatively accurate in the 2D limit $\hbar\omega_z\gg\mu$, yet still provide 
qualitatively correct predictions in the quasi2D and Thomas-Fermi regimes. 

Averaging Eq.~\eqref{EQHeisenberg3D} over the axial degree of freedom, we find 
\begin{equation}\label{EQHeisenberg2D}
 i\hbar\partial_t\hat{\psi}=\left[\frac{-\hbar^2}{2m}\nabla_{\bs{\rho}}^2+
V_{\rm int}^{\rm 2D}(\bs{\rho})\right]\hat{\psi}
\end{equation}
where $V_{\rm int}^{\rm 2D}(\bs{\rho})=
\frac{1}{2}\int
d^2\bs{\rho}'\hat{\psi}^\dagger(\bs{\rho}') U_{\rm 2D}(\bs{\rho}-\bs{\rho}')
\hat{\psi}(\bs{\rho}')$. 
The two body interactions are now described by the potential $U_{\rm 2D}(\bs{\rho})$ 
which includes an interaction with the mean density along the $z$-axis, and is 
given by
\begin{equation}\label{EQU2D}
 U_{\rm 2D}(\bs{\rho})=
\frac{D}{3\sqrt{2\pi}l_z}\int \frac{d^2\mathbf{k}}{(2\pi)^2}
e^{-i\mathbf{k}\cdot\boldsymbol{\rho}}\left[2-3\sqrt{\pi}h\left(\frac{kl_z}{\sqrt{2}}\right)\right]
\end{equation}
where $h(x)=xe^{x^2}{\rm erfc}\left(x\right)$. This interaction potential asymptotes toward 
$U_{\rm 2D}(\bs{\rho})\rightarrow D\pi^{5/2}/(12 \sqrt{2})\left(l_z/\rho\right)^3$ when 
$\rho\gg l_z$, but then becomes negative (attractive interactions) when $\rho\lesssim l_z$. 
This attractive component is directly due to the finite density of particles along the 
$z$ axis. 

At zero temperature, we assume the single particle reduced density matrix has a macroscopically 
occupied eigenstate, which can be treated as a classical field (in an ensemble in which the phase-symmetry 
of $\hat{\psi}$ has been broken). Under this assumption, the field operator can be split up into 
\begin{equation}
 \hat{\psi}(\bs{\rho})=\phi(\bs{\rho})+\hat{\delta}(\bs{\rho})
\end{equation}
where $\phi$ describes the macroscopically occupied state (the condensate wave function) and 
$\hat{\delta}$ describes all other states (which are assumed to have small occupation numbers). 
As a first order approximation, only the particles in the condensate are considered, and 
Eq.~\eqref{EQHeisenberg2D} reduces to the familiar dipolar Gross-Pitaevskii equation,
\begin{equation}\label{EQDipolarGPE}
 i\hbar\partial_t\phi(\bs{\rho})=\left[\frac{-\hbar^2}{2m}\nabla_{\bs{\rho}}^2+
\int d^2\bs{\rho}'n_0(\bs{\rho}') U_{\rm 2D}(\bs{\rho}-\bs{\rho}')\right]\phi(\bs{\rho})
\end{equation}
where $n_0=\phi^*\phi$ is the condensate density. 

The equilibrium solution to Eq.~\eqref{EQDipolarGPE} is $\phi=\sqrt{n_0}e^{-i\mu t/\hbar}$ where 
the chemical potential is given by 
\begin{equation}
 \mu=\sqrt{\frac{2}{\pi}}\frac{Dn_0}{3l_z},
\end{equation}
and $n_0$ denotes the projected 2D density (that is, the density of all particles projected onto the 
$x$-$y$ plane). 

The noncondensate particles, which occupy states contained in $\hat{\delta}$ can be solved for, 
within the Bogoliubov approximation, by substituting $\hat{\psi}=\sqrt{n_0}e^{-i\mu t/\hbar}+\hat{\delta}$ 
into Eq.~\eqref{EQHeisenberg2D}. Ignoring terms of order $\hat{\delta}^2$ and higher (which describe interactions 
between noncondensate particles), and employing the Bogoliubov transformation
\begin{equation}
 \hat{\delta}=e^{-i\mu t/\hbar}\int d^2\mathbf{k}\left[u_ke^{-iE_kt/\hbar}
\hat{a}_k^{\phantom{\dagger}}+v_k^*e^{iE_kt/\hbar}\hat{a}_k^\dagger\right]
\end{equation}
we obtain the following coupled equations for the quasi-particle amplitudes,
\begin{align}
\label{EQbdg1equib}
 E_{k}u_k=E_k^f u_{k}+n_0\left(u_{k}+v_{k}\right)\mathcal{F}_k\left[U_{\rm 2D}\right]
\\
\label{EQbdg2equib}
-E_{k}v_k=E_k^f v_{k}+n_0\left(u_{k}+v_{k}\right)\mathcal{F}_k\left[U_{\rm 2D}\right]
\end{align}
where $E_k^f=\frac{\hbar^2k^2}{2m}$ is the energy of a free particle and $\mathcal{F}_k\left[U_{\rm 2D}\right]=\frac{D}{3\sqrt{2\pi}l_z}
\left[2-3\sqrt{\pi}h\left(\frac{kl_z}{\sqrt{2}}\right)\right]$ is the Fourier transform of the two-body interaction 
in Eq.~\eqref{EQU2D}. Equations~\eqref{EQbdg1equib} and~\eqref{EQbdg2equib} can be solved, to give
\begin{align}
 u_{{k}}=\frac{1}{4\pi}\frac{E_k^f+E_{k}}{\sqrt{E_{k}E_k^f}},\label{EQuk1}\\
 v_{{k}}=\frac{1}{4\pi}\frac{E_k^f-E_{k}}{\sqrt{E_{k}E_k^f}},\label{EQvk1}
\end{align}
and the excitation spectrum is given by
\begin{equation}\label{EQspectrum}
 E_{\vec{k}}^2=E_k^f\left[E_k^f+2n_0\mathcal{F}_k\left[U_{\rm 2D}\right]\right].
\end{equation}
Using these solutions to the Bogoliubov-de Gennes equations, we can investigate 
equilibrium correlation functions of the system, which is the aim of this paper.

\section{Excitation spectrum and stability criteria}\label{SECexcitation}

A particularly interesting feature of this system is the emergence of a roton in the excitation 
spectrum, Eq.~\eqref{EQspectrum}. 
This spectrum can be rewritten by factorising out 
the chemical potential and defining $\mathbf{q}=\xi\mathbf{k}$ where 
$\xi=\hbar/\sqrt{m\mu}$ defines a healing length (or correlation length) for the fluid. This 
yields
\begin{equation}
 E_k=\mu q\sqrt{\frac{q^2}{4}+1-\frac{3\sqrt{\pi}}{2}h\left(\frac{q\sigma}{\sqrt{2}}\right)},
\end{equation}
where $\sigma=l_z/\xi$ is the ratio of confinement length to healing length. So we find that 
the strength of the roton feature in this system depends only on one parameter; $\sigma$. Using 
the definition of $\xi$ and $\mu$ we can rewrite 
\begin{equation}\label{EQsigma}
 \sigma=\frac{l_z}{\xi}=2^{5/4}\pi^{1/4}\sqrt{a_{dd}n_0l_z}
\end{equation}
where $a_{dd}=mD/(12\pi\hbar^2)$ is the dipolar length scale~\cite{DipolarBosonsReview}. 
From Eq.~\eqref{EQsigma} 
we see that the prominence of the roton in the spectrum depends on the number of particles within 
one dipole-length-scale $\sim\sqrt{n}a_{dd}$, multiplied by the number of particles within 
one confinement width $\sim\sqrt{n}l_z$.

\begin{figure}
\centering
\includegraphics[width=7cm]{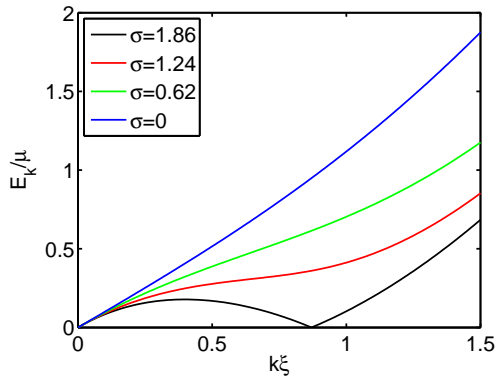}
\caption{The excitation spectrum for quasi2D dipolar Bose gas given by Eq.~\eqref{EQspectrum}.}
\label{FIGSpectrum}
\end{figure}

We find that the short-range attractive component of $U_{\rm 2D}$ renders the system unstable 
beyond a critical value of $\sigma$ which we numerically determine to be 
$\sigma_c\approx 1.8562$. For $\sigma>\sigma_c$ there exists complex 
values of $E_k$, indicating a dynamical instability. 

Interestingly, the excitation spectrum at $\sigma=0$ is exactly equivalent to a system 
with contact interactions.

\section{Noncondensate density}\label{SECnoncondensate}

\begin{figure}
\centering
\includegraphics[width=7cm]{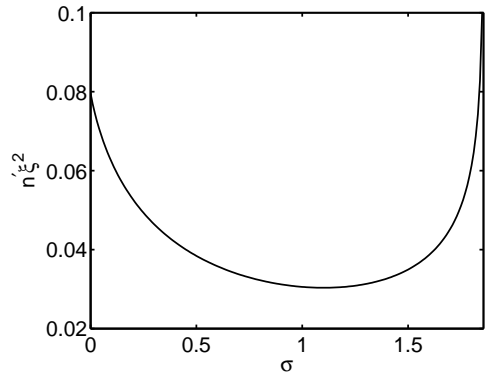}
\caption{The noncondensate density for a quasi2D dipolar Bose gas as a function of $\sigma$. 
The plot is generated from a numerical solution to Eq.~\eqref{EQnoncondensatedensity}.}
\label{FIGNoncondensateDensity}
\end{figure}

Calculating the noncondensate density provides an important retrospective analysis of 
how accurate the assumptions of Bogoliubov theory are. The noncondensate 
density is given by,
\begin{equation}
 n'=\langle\hat{\delta}^\dagger\hat{\delta}\rangle=\frac{1}{(4\pi)^2}\int d^2\mathbf{k} \;\left[\frac{E_k^f}{E_k}
+\frac{E_k}{E_k^f}-2\right],
\end{equation}
and if we use $\mathbf{q}=\xi\mathbf{k}$, and switch to polar coordinates, we find, 
\begin{equation}\label{EQnoncondensatedensity}
 n'=\frac{1}{8\pi\xi^2}\int_0^\infty dq\;\left[
\frac{q^2+2-3\sqrt{\pi}h\left(\frac{q\sigma}{\sqrt{2}}\right)}{\sqrt{\frac{q^2}{4}+1-
\frac{3\sqrt{\pi}}{2}h\left(\frac{q\sigma}{\sqrt{2}}\right)}}-2q
\right].
\end{equation}
A necessary criteria to ensure our formalism is accurate, is that the noncondensate density 
be much less than the condensate density, $n'/n_0\ll1$. Immediately from Eq.~\eqref{EQnoncondensatedensity} 
we see that this criteria will be satisfied if $\xi^2n_0$ is sufficiently large. That is, the 
number of particles within one square healing length, needs to be large. Of course the 
ratio also depends on the value of the integral in Eq.~\eqref{EQnoncondensatedensity}, which depends 
on $\sigma$. We numerically solved this integral and the result is shown in Fig.~\ref{FIGNoncondensateDensity}. 
We find the noncondensate density $n'=1/4\pi$ when $\sigma=0$, and initially decreases as $\sigma$ increases. 
However, when $\sigma$ becomes larger than $\approx1.5$, the noncondensate density begins to rapidly diverge; 
$n'\rightarrow\infty$ as $\sigma\rightarrow\sigma_c$. This divergence is due to a rapid increase in the 
population of the roton modes. This fact will become clear after examining the momentum distribution of the system.

\section{Momentum distribution}\label{SECmomentum}

We calculate the momentum distribution of the system as a means of analysing the 
occupation number of the roton modes. In general, this can be found via the Fourier transform 
of the one-body density operator,
\begin{equation}
 \eta(\mathbf{k})=\frac{1}{A(2\pi)^2}\int d^2\bs{\rho}\int d^2\bs{\rho}' \langle\hat{\psi}^\dagger(\bs{\rho})
\hat{\psi}(\bs{\rho}')\rangle e^{i\mathbf{k}\cdot(\bs{\rho}-\bs{\rho}')}
\end{equation}
where $A$ is the total area of the 2D system. Within our current formalism, this reduces down to
\begin{equation}\label{EQmomentumdistribution}
 \eta(\mathbf{k})=n_0\delta(\mathbf{k})+\frac{1}{(4\pi)^2}\left[
\frac{E_k^f}{E_k}
+\frac{E_k}{E_k^f}-2
\right].
\end{equation}
The first term in Eq.~\eqref{EQmomentumdistribution} is simply the condensate mode, and the 
second term constitutes the so-called quantum depletion caused by the interactions. This 
quantum depletion is zero, only in the case where there are no interactions 
in the system and $E_k=E_k^f$.

\begin{figure}
\centering
\includegraphics[width=7cm]{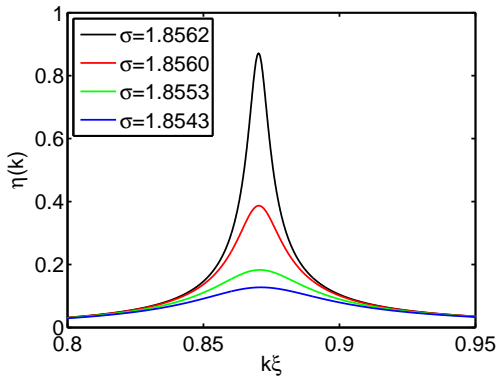}
\caption{The momentum distribution, Eq.~\eqref{EQmomentumdistribution}, for a quasi2D dipolar Bose gas for different values of $\sigma$. 
The distribution shows a remarkable sensitivity to $\sigma$ as it approaches $\sigma_c$.}
\label{FIGMomentumDist}
\end{figure}

We analyse the asymptotic behaviour of $\eta(\mathbf{k})$, and find that for $k\xi\ll 1$, 
the distribution depends weakly on the value of $\sigma$;
\begin{equation}
 \eta(\mathbf{k})\approx n_0\delta(\mathbf{k})+\frac{1}{(4\pi)^2}\left(\frac{2}{k\xi}-2-
\frac{3\sqrt{\pi}\sigma}{2\sqrt{2}}\right).
\end{equation}
However, in the region $0.75\lesssim k\xi\lesssim 1$ where the roton appears in the spectrum, 
the momentum distribution shows a remarkable sensitivity to the value of $\sigma$, as shown 
in Fig.~\ref{FIGMomentumDist}. As $\sigma$ approaches $\sigma_c$, we see a dramatic increase 
in the population of these roton modes. This behaviour is similar to that shown in a quasi-1D 
geometry for laser induced dipoles~\cite{ODellDepletion}. This increase in population explains the rapid 
divergence of the noncondensate density seen in Fig.~\ref{FIGNoncondensateDensity}. 
As the system approaches the instability, macroscopic occupations of these roton modes begin to exist, 
resulting in the systems eventual collapse.

\section{Phase coherence}\label{SECphase}

\begin{figure}
\centering
\includegraphics[width=7cm]{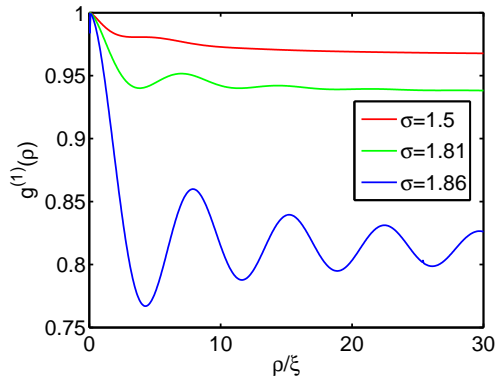}
\caption{The first order correlation function, $g^{(1)}$ of Eq.~\eqref{EQg1}, for a quasi2D dipolar Bose gas for different values of $\sigma$. 
The phase coherence shows an unusual oscillatory behaviour as $\sigma\rightarrow\sigma_c$.}
\label{FIGg1}
\end{figure}

In this section, we calculate the first order correlation function for the system
\begin{equation}
 g^{(1)}(\bs{\rho}-\bs{\rho}')=\frac{\langle\hat{\psi}^\dagger(\bs{\rho})
\hat{\psi}(\bs{\rho}')\rangle}{n}
\end{equation}
where $n=n_0+n'$ is the total density of the system. This quantity tells us 
how correlated the phase of a given particles wavefunction at some point $\bs{\rho}$ is 
with the phase of that same single-particle-wavefunction at another point $\bs{\rho}'$. 
A Bose-Einstein condensate is said to exist when $g^{(1)}(\bs{\rho})$ decays to some finite 
(nonzero) value as $\rho\rightarrow\infty$. A pure Bose-Einstein condensate would imply 
$g^{(1)}(\bs{\rho})=1$ everywhere, exactly as in Glauber's original definition of a coherent state. 
Interactions in the system destroy the coherence, in our case this is given by
\begin{equation}
g^{(1)}(\bs{\rho})=\frac{n_0}{n}+\frac{1}{n(4\pi)^2}\int d^2\mathbf{k}\;e^{i\mathbf{k}\cdot\bs{\rho}}
\left[\frac{E_k^f}{E_k}
+\frac{E_k}{E_k^f}-2\right].
\end{equation}
Again we switch to the dimensionless variable $\mathbf{q}=\xi\mathbf{k}$, and convert to 
polar coordinates, to find
\begin{align}\label{EQg1}
 g^{(1)}&(\bs{\rho})=\frac{n_0}{n}+\frac{1}{8\pi n\xi^2}\times\nonumber\\
&\int_0^\infty \!\!\! dq\;
J_0\left(\frac{q\rho}{\xi}\right)\left[
\frac{q^2+2-3\sqrt{\pi}h\left(\frac{q\sigma}{\sqrt{2}}\right)}{\sqrt{\frac{q^2}{4}+1-
\frac{3\sqrt{\pi}}{2}h\left(\frac{q\sigma}{\sqrt{2}}\right)}}-2q
\right]
\end{align}
where $J_0$ is a Bessel function. For the case $\sigma=0$, where the spectrum 
resembles that of contact interactions, the phase coherence is given by,
\begin{equation}
 g^{(1)}(\bs{\rho})=\frac{n_0}{n}+\frac{1}{2\pi n\xi^2}I_1(\rho/\xi)K_1(\rho/\xi)
\end{equation}
where $I_1$ and $K_1$ are Bessel functions. 
For $\sigma\neq0$, we solve the integral in Eq.~\eqref{EQg1} numerically, and 
plot the results in Fig.~\ref{FIGg1}. Interestingly, the correlation function 
displays an oscillatory nature for values of $\sigma$ close to $\sigma_c$. 
This oscillation implies partial-collapse/revival type behaviour, as if the system 
is attempting to fragment into smaller pieces.

\section{Density fluctuations}\label{SECdensity}

\subsection{Two-body correlations}

Two-body, density-density correlations yield important information with respect to the 
linear-response of the system to external perturbations. The quantity of interest to us, 
is the so-called $g^{(2)}(\bs{\rho},\bs{\rho}')$ function, which quantifies the 
conditional probability of detecting a particle at $\bs{\rho}$, given that one 
has been detected at $\bs{\rho}'$. This quantity is defined as 
\begin{equation}
 g^{(2)}(\bs{\rho}-\bs{\rho}')=\frac{\langle
\hat{\psi}^\dagger(\bs{\rho})\hat{\psi}^\dagger(\bs{\rho}')
\hat{\psi}(\bs{\rho}')\hat{\psi}(\bs{\rho})
\rangle}{n^2}
\end{equation}
which in our case reduces down to 
\begin{equation}
 g^{(2)}(\bs{\rho}-\bs{\rho}')=1+\frac{1}{(2\pi)^2n}\int d^2\mathbf{k}\;
e^{i\mathbf{k}\cdot(\bs{\rho}-\bs{\rho}')}\left(\frac{E_k^f}{E_k}-1\right).
\end{equation}
Switching to the dimensionless variable $\mathbf{q}=\mathbf{k}\xi$, and converting to 
polar coordinates, we find
\begin{align}
 g^{(2)}&(\bs{\rho})=1+\frac{1}{2\pi\xi^2n}\times\nonumber\\
&\int_0^\infty dq\;J_0\left(\frac{q\rho}{\xi}\right)
\left[
\frac{q}{2\sqrt{\frac{q^2}{4}+1-\frac{3\sqrt{\pi}}{2}h\left(\frac{q\sigma}{\sqrt{2}}\right)}}-1
\right].\label{EQg2}
\end{align}
The integral in Eq.~\eqref{EQg2} converges for all $\rho\neq0$. When $\rho=0$ however, the integral has 
an ultraviolet divergence. The physical reason for this divergence is well known, and is related to 
the incorrect expression for the short-range interaction in $U_{\rm2D}$. Physically, the interaction 
potential $U_{\rm 2D}$ should include a short-range van der Waals type interaction, which would introduce 
a short-wavelength cut-off (a finite upper-limit) to the integral of Eq.~\eqref{EQg2}. This ultraviolet 
cut-off does not significantly alter the behaviour for $\rho\neq0$, and the local correlation at 
$\rho=0$ is only logarithmically dependent on this cut-off.

\begin{figure}
\centering
\includegraphics[width=8cm]{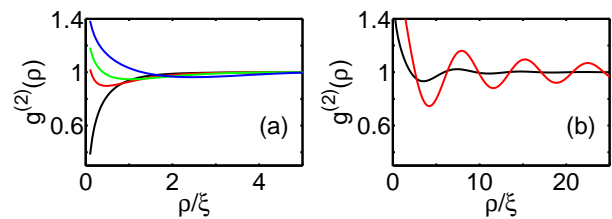}
\caption{The second order, density-density correlation function, $g^{(2)}$ of Eq.~\eqref{EQg2}, for a quasi-2D dipolar Bose gas 
for different values of $\sigma$. The different colors correspond to different values of $\sigma$, in (a) 
the black line is $\sigma=0$, red line is $\sigma=0.38$, green line is $\sigma=0.75$, and blue line is 
$\sigma=1.5$. In (b), the black line is $\sigma=1.81$, and red line is $\sigma=1.8562$.}
\label{FIGg2}
\end{figure}

The numerical solutions to the integral in Eq.~\eqref{EQg2} are shown in Fig.~\ref{FIGg2}. 
We see that, when $\sigma=0$, the case is equivalent to repulsive contact interactions, where 
the repulsion between particles causes a decrease in the probability of detecting two particles 
close to one another i.e. $g^{(2)}<1$. As $\sigma$ increases, particles begin to feel the effect of the third dimension, 
that is, the attractive core in the interparticle potential $U_{\rm 2D}$. This attraction causes 
an increase in the local value of $g^{(2)}>1$. The most intriguing behaviour in $g^{(2)}$ appears near the 
critical point, $\sigma\approx\sigma_c$. Near this point we see $g^{(2)}(\rho)$ dip below unity, as one would 
expect for repulsive long range interactions, but then oscillate between local maxima/minima which 
are above and below unity respectively. The first and highest peak occurs at $\rho=0$, and is caused by 
the remnant attractive component of the 3D interactions. The second peak, occurs at a distance 
$r_{\rm max}\approx7.92\xi$, and can be loosely construed as a preferred separation distance between particles 
(or possibly clumps of particles). 
Subsequent peaks occur at approximately integer multiples of $r_{\rm max}$. 
This behaviour seems qualitatively similar to what one might expect 
from a fluid which is undergoing a transition to a solid state, where long range diagonal order would 
be present. Because the translational symmetry of the system remains unbroken, the diagonal order 
becomes off-diagonal order in the $g^{(2)}(\bs{\rho},\bs{\rho}')$ function. The order which we find in 
this case, however, is not long range, and decays proportional to $1/|\bs{\rho}-\bs{\rho}'|^{1/2}$.

\subsection{Higher order correlations}

To further test the hypothesis that the system is attempting to form an ordered state, we also study 
higher order correlations in the system, three-body, four-body, etc. In general the $N$th order 
density correlation is given by,
\begin{equation}
 g^{(N)}(\bs{\rho}_1,\ldots,\bs{\rho}_N)=
\frac{\langle
\hat{\psi}^\dagger(\bs{\rho}_1)\ldots\hat{\psi}^\dagger(\bs{\rho}_N)
\hat{\psi}(\bs{\rho}_N)\ldots\hat{\psi}(\bs{\rho}_1)
\rangle}{n^N},
\end{equation}
this can be interpreted as a conditional probability of detecting a particle at 
position $\bs{\rho}_1$, given the detection of $N-1$ particles at positions 
$\bs{\rho}_2,\ldots,\bs{\rho_N}$. Within our current theory, this function can be 
simplified down to 
\begin{align}
 g^{(N)}&=1+\frac{1}{2\pi\xi^2n}\sum_{\langle i,j\rangle}\int_0^\infty dq\;
J_0\left(\frac{q|\bs{\rho}_i-\bs{\rho}_j|}{\xi}\right)\times\nonumber\\
&\left[
\frac{q}{2\sqrt{\frac{q^2}{4}+1-\frac{3\sqrt{\pi}}{2}h\left(\frac{q\sigma}{\sqrt{2}}\right)}}-1
\right]
\end{align}
where the notation; $\sum_{\langle i,j\rangle}=\sum_{j=2}^N\sum_{i=1}^j$ is the summation over 
all pairs of indices, and the arguments of $g^{(N)}$ have been omitted for notational convenience. 

\begin{figure*}
\centering
\includegraphics[width=16cm]{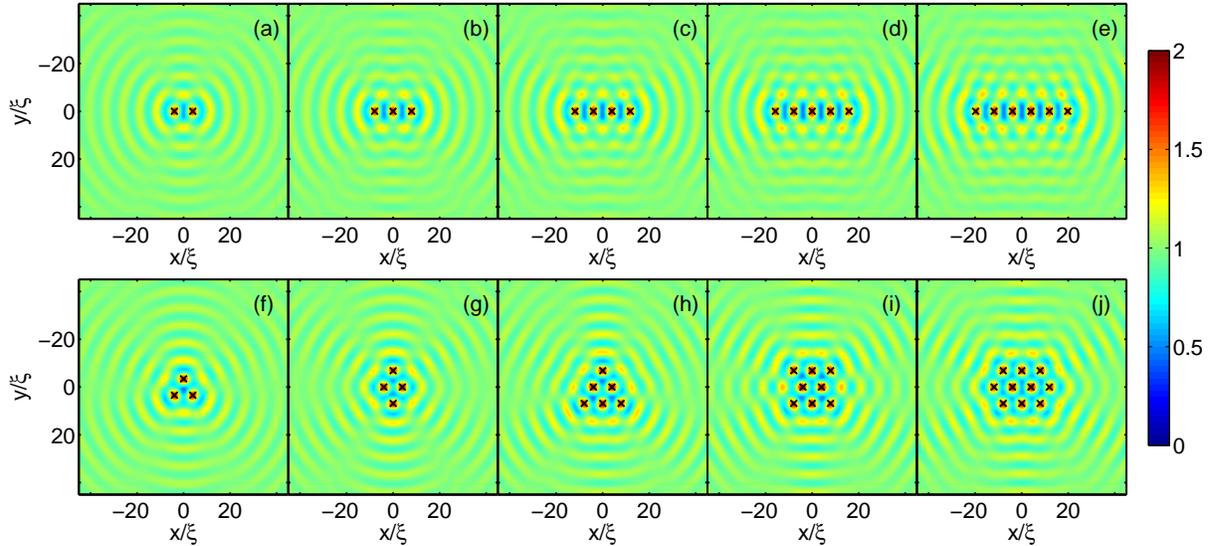}
\caption{Higher order density fluctuations for $\sigma=1.8562$. In (a)--(j) we plot 
$g^{(N)}(\bs{\rho}_1,\ldots,\bs{\rho}_{N-1},\bs{\rho})-g^{(N-1)}(\bs{\rho}_1,\ldots,\bs{\rho}_{N-1})$,
for a variety of different values of $N$, aswell as different choices of $\bs{\rho}_1,\ldots,\bs{\rho}_{N-1}$. 
The black crosses show the positions of $\bs{\rho}_1,\ldots,\bs{\rho}_{N-1}$, which are chosen 
according to the procedure described in the text. From the number of black crosses one can deduce the 
value of $N$ in each subfigure. We note the intriguing emergence of hexagonal order in 
the density distributions. }
\label{FIGgN}
\end{figure*}

To visualise this high dimensional observable, we choose fixed values for 
$\bs{\rho}_1$, up to $\bs{\rho}_{N-1}$ in the following way. The first particle is 
detected at some arbitrary point. This choice is arbitrary, as there is complete 
translational symmetry in the system at this point. Subsequent particles are then chosen at the 
most likely points given by the behaviour of $g^{(2)}(\bs{\rho}_1,\bs{\rho}_2)$, 
$g^{(3)}(\bs{\rho}_1,\bs{\rho}_2,\bs{\rho}_3)$, etc, with the proviso that these 
new points are spatially separated from the previous points. For instance, 
consider Fig~\ref{FIGgN}(a), where we plot $g^{(3)}(\bs{\rho}_1,\bs{\rho}_2,\bs{\rho})$. We 
begin by (arbitrarily) setting $\bs{\rho}_1=(-r_{\rm max}/2,0)$. We know then, from our calculation 
of $g^{(2)}$ in the previous subsection, that there will be an increased probability of detecting 
a second particle somewhere on the ring  centred at $\bs{\rho}_1$, and with a radius of $r_{\rm max}$ 
[see the red line in Fig.~\ref{FIGg2}(b)]. 
So we choose to place the second particle at the point  $\bs{\rho}_2=(r_{\rm max}/2,0)$, such that 
the two particles are separated by a distance of exactly $r_{\rm max}$. We then plot the 
function $g^{(3)}(\bs{\rho}_1,\bs{\rho}_2,\bs{\rho})-g^{(2)}(\bs{\rho}_1,\bs{\rho}_2)$ for 
this choice of $\bs{\rho}_1$ and $\bs{\rho}_2$, shown in Fig.~\ref{FIGgN}(a) (note that 
we have subtracted off 
the lower-order $g^{(2)}(\bs{\rho}_1,\bs{\rho}_2)$ so that the asymptotic value of the plotted 
function $\rightarrow1$ as $\rho\rightarrow\infty$). The black crosses in the figure indicate the 
positions of $\bs{\rho}_1$ and $\bs{\rho}_2$. As was the case, with $g^{(2)}$, we again see 
a series of maxima/minima which are above and below zero respectively. The maxima along the $x$-axis 
are placed at integer multiples of $(r_{\rm max}/2,0)$. However, along the $y$-axis the first maxima is 
at approximately $(0,\pm r_{\rm max}\sqrt{3}/2)$, consistent with the formation of a triangular lattice 
(the densest packing arrangement of spheres in 2D).

Continuing on in this fashion, we construct higher order correlation functions, and visualise them in
this way. The results are shown in Fig.~\ref{FIGgN} for a variety of different choices of 
$\bs{\rho}_1,\ldots,\bs{\rho}_{N-1}$. The results suggest the formation of localised order 
consistent with a triangular lattice of high-density droplets, separated by a distance of 
approximately $r_{\rm max}$. The value of $\sigma$ is very close to the critical value 
$\sigma_c$, where the theory becomes invalid. For this reason, we proceed with a 
degree of scepticism, and only claim that, although the quantitative aspect of these 
predictions is questionable, the qualitative aspect seems very reasonable. Specifically, 
the results give us clues as to the ground state of the system past the phonon-instability. 
This problem was studied numerically in Ref.~\cite{NathPhononInstabilityTwoDimDipoles}, where 
they conluded that, beyond the critical point, the system forms a soliton gas. These conclusions 
were reached by simulating the dipolar Gross-Pitaevskii equation, with an instantaneous quench in the 
dipolar coupling constant such that $\sigma>\sigma_c$. We conject that it may be possible to, instead, 
form a state in which the solitons spatial distribution is ordered into a triangular lattice, 
somewhat like a soliton solid. This state could be reached if the dipolar coupling constant is 
increased adiabatically, or at least very slowly, beyond $\sigma_c$. 

\begin{figure}
\centering
\includegraphics[width=7cm]{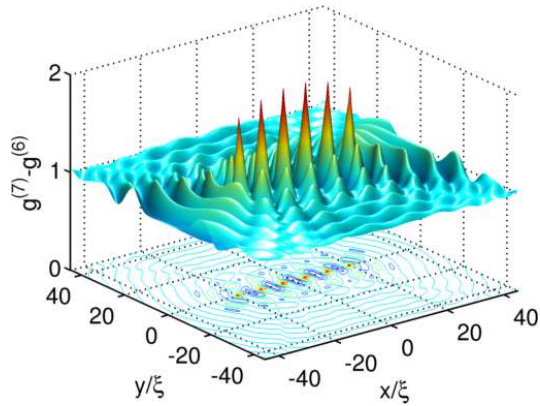}
\caption{Higher order density fluctuations for $\sigma=1.8562$, as in Fig.~\ref{FIGgN}(e) 
shown as a surface plot with contour lines.}
\label{FIGgNfancy}
\end{figure}

\section{Conclusions and Discussion}\label{SECconclusions}

In this article, we have formulated a simple theoretical model for predicting the 
behaviour of quasi-2D dipolar Bosons close to zero temperature. We solved the 
model in the case of a uniform system in equilibrium, and studied the 
excitation spectrum, the momentum distribution, the phase fluctuations and the 
density fluctuations. Regarding the excitation spectrum, roton modes form in the vicinity 
of $k\approx0.87/\xi$ due to a competition between the trapping potential and the attractive interaction 
in that direction. We found 
a single parameter, $\sigma$ defined in Eq.~\eqref{EQsigma}, which determines 
the strength of the roton, and the point of 
collapse. For the momentum distribution, we found it decays as $1/\bf{k}$ when $k\xi\ll1$. 
However, as the system approaches the critical point, the occupation of the roton modes begin 
to increase rapidly. Eventually these roton modes become macroscopically occupied 
resulting the eventual collapse of the system. The level of phase fluctuations which occur in the system is 
comparable to the fraction of noncondensate particles in the system. This initially decreases 
with increasing $\sigma$, but then rapidly diverges as the roton modes become highly occupied. 
The density fluctuations display a variety of different behaviour due to competition between the 
interatomic repulsion at large distances versus the short range attraction at short distances (due to the 
mean field of particles in the trapped direction). The interplay of these different length scales 
manifests as the behaviour shown in Fig.~\ref{FIGg2}(a). As $\sigma$ approaches the critical point, 
the density fluctuations of $g^{(2)}$ begin to show an anomalous oscillatory behaviour, with maxima and minima 
above and below unity, respectively. We construe this oscillatory behaviour as a preferred 
separation distance between particles or clumps of particles. We further pursued this idea by 
calculating higher order density fluctuations, shown in Fig.~\ref{FIGgN}. 
The coexistence of crystalline order and superfluidity (supersolidity) has been a subject of 
intense recent debate~\cite{ChanSuperSolid}. Although what we observe here is not true 
crystalline order, since it is not long range, it does provide an intriguing possibility 
for the manifestation of a supersolid state.

\subsection{Relation to experiments}
We conclude with some brief remarks on the possibility of experimental observation 
of our results. To date, dipolar Bose-Einstein condensates have been 
achieved with Chromium~\cite{ChromiumDipolarEffects2}, Dysprosium~\cite{DysprosiumCondensate}, 
and Erbium~\cite{ErbiumCondensate}, which have dipole length 
scales of approximately 0.85nm, 7.7nm, and 3.9nm respectively. We assume 
2D densities of around $10^{13}{\rm m}^{-2}$ to $10^{15}{\rm m}^{-2}$. 
These estimates are approximately based on what has 
been achievable in traditional cold atom systems such as Rubidium~\cite{DalibardBKT}. 
We see then that the if the confinement width, $l_z$, is on the order of 2-3$\mu$m 
then $\sigma$ will lie in the stable regime, and by tuning $l_z$ one should be able 
to observe the roton facilitated collapse.

A viable alternative to ultra-cold gases is in systems of strongly correlated electrons, 
such as the exciton gases found in certain semiconducting heterostructures~\cite{ExcitonBEC}. 
Such systems have been studied using a path-integral Monte-Carlo method in Ref.~\cite{MonteCarloExcitons}.

\section{Acknowledgements}

AGS wishes to thank Robert Ecke and Avadh Saxena for motivating discussions. 
Both authors gratefully acknowledge the support from 
LANL, which is operated by LANS, LLC for the NNSA 
of the US DOE under Contract No. DE-AC52-06NA25396. 
One author (CT) acknowledges support from the 
ASC Program and the other author (AS) from the LDRD program.


\begin{thebibliography}{99}

\bibitem{FeshbachReview} Cheng Hin, Rudolf Grimm, Paul Julienne, and Eite Tiesinga, Rev. Mod. Phys. 
{\bf 82}, 1225 (2010).
\bibitem{KetterleFeshbachDemonstration} S. Inouye, {\it et. al.}, Nature {\bf 392}, 151 (1998).

\bibitem{ChromiumCondensate} A. Griesmaier, J. Werner, S. Hensler, J. Stuhler, and T. Pfau, 
Phys. Rev. Lett. {\bf 94}, 160401 (2005).
\bibitem{ChromiumDipolarEffects1} Thierry Lahaye, {\it et. al.}, Nature {\bf 448}, 672 (2007).
\bibitem{ChromiumDipolarEffects2} T. Koch, T. Lahaye, J. Metz, B. Fr\"olich, A. Griesmaier, and T. Pfau,
Nature Physics {\bf 4}, 218 (2008).

\bibitem{DysprosiumCondensate} Mingwu Lu, Nathaniel Q. Burdick, Seo Ho Youn, and Benjamin L. Lev, 
Phys. Rev. Lett. {\bf 107}, 190401 (2011).

\bibitem{ErbiumCondensate} K. Aikawa, A. Frisch, M. Mark, S. Baier, A. Rietzler, R. Grimm, and F. Ferlaino, 
Phys. Rev. Lett. {\bf 108}, 210401 (2012).

\bibitem{DipolarBosonsReview} T. Lahaye, C. Menotti, L. Santos, M. Lewenstein, ant T. Pfau, 
Rep. Prog. Phys. {\bf 72}, 126401 (2009).

\bibitem{DebbieJinKRbMolecules} K.-K. Ni, {\it et. al.}, Science {\bf 322}, 231 (2008).
\bibitem{RbCsMolecules} Tetsu Takekoshi, {\it et. al.}, Phys. Rev. A {\bf 85}, 032506 (2012).


\bibitem{SpielmanSyntheticPartialWaves} R. A. Williams, {\it et. al.}, Science {\bf 335}, 314 (2012).




\bibitem{FischerTwoDimDipoleStability} Uwe R. Fischer, Phys. Rev. A {\bf 73}, 031602(R) (2006).

 
\bibitem{BissettNormalStability} R. N. Bisset, D. Baillie, and P. B. Blakie, Phys. Rev. A {\bf 83}, 061602(R) (2011).

\bibitem{ChineseNormalFermiGas} J.-N. Zhang and S. Yi, Phys. Rev. A {\bf 81}, 033617 (2010).
\bibitem{BaillieNormalFermiGas} D. Baillie and P. B. Blakie, Phys. Rev. A {\bf 82}, 033605 (2010).

\bibitem{DipolarCooperPairs} M. A. Baranov, M. S. Mar'enko, Val. S. Rychkov, and G. V. Shlyapnikov, 
Phys. Rev. A {\bf 66}, 013606 (2002).
\bibitem{DipolarFermionicSuperfluid} M. A. Baranov, L. Dobrek, and M. Lewenstein, Phys. Rev. Lett. {\bf 92}, 250403 (2004).
\bibitem{DipolarFermionsQuantumHall} M. A. Baranov, Klaus Osterloh, and M. Lewenstein, Phys. Rev. Lett. {\bf 94}, 070404 (2005).

\bibitem{BlakieDipolarGPENumerics} P. B. Blakie, C. Ticknor, A. S. Bradley, A. M. Martin, M. J. Davis, and Y. Kawaguchi,
Phys. Rev. E {\bf 80}, 016703 (2009).
\bibitem{OtherEfficientDipolarNumerics} Weizhu Bao, Yongyong Cai, and Hanquan Wang, Journal of Computational Physics 
{\bf 229}, 7874 (2010).

\bibitem{RonenDipolarCylindricalTrapBogoliubovModes} Shai Ronen, Daniele C. E. Bortolotti, and John L. Bohn, 
Phys. Rev. A {\bf 74}, 013623 (2006).

\bibitem{SantosRoton} L. Santos, G. V. Shlyapnikov, and M. Lewenstein, Phys. Rev. Lett. {\bf 90}, 250403 (2003).

\bibitem{WilsonManifestationsOfDipolarRotonMode} Ryan M. Wilson, Shai Ronen, John L. Bohn and Han Pu, Phys. Rev. Lett. 
{\bf 100}, 245302 (2008).

\bibitem{PalevskyNeutronScatteringHelium} H. Palevsky, K. Otnes, and K. E. Larsson, Phys. Rev. {\bf 112}, 11 (1958). 

\bibitem{WilsonDipolarCriticalVelocity} Ryan M. Wilson, Shai Ronen, and John L. Bohn, Phys. Rev. Lett. {\bf 104}, 
094501 (2010).
\bibitem{TicknorAnisotropicSuperfluidity} Christopher Ticknor, Ryan M. Wilson, and John L. Bohn, 
Phys. Rev. Lett. {\bf 106}, 065301 (2011).

\bibitem{YiPuVorticesDipolarBosons} S. Yi, and H. Pu, Phys. Rev. A {\bf 73}, 061602(R) (2006).
\bibitem{WilsonVortexDipolarBosons} Ryan M. Wilson, Shai Ronen, and John L. Bohn, Phys. Rev. A {\bf 79}, 013621 (2009).

\bibitem{AnisotropicSolitonDipolarCondensates} I. Tikhonenkov, B. A. Malomed, and A. Vardi, Phys. Rev. Lett. {\bf 100},
090406 (2008).
\bibitem{TwoDimBrightSolitionsDipolarCondensates} Patrick K\"oberle, Damir Zajec, G\"unter Wunner, and Boris A. Malomed, 
Phys. Rev. A {\bf 85}, 023630 (2012)

\bibitem{AstrakharchikMeanFieldLimitations} G. E. Astrakharchik, J. Boronat, J. Casulleras, 
I. L. Kurbakov, and Y. E. Lozovik, Phys. Rev. A {\bf 75},
063630 (2007).
\bibitem{JainMonteCarloDipoles} Piyush Jain, Fabio Cinti, and Massimo Boninsegni, Phys. Rev. B {\bf 84}, 014534 (2011).

\bibitem{KaliaVashishta} R. K. Kalia and P. Vashishta, J. Phys. C: Solid State Phys. {\bf 14}, L643-L648 (1981).
\bibitem{ZollerStronglyCorrelatedDipoles} H. P. B\"uchler, {\it et. al.}, Phys. Rev. Lett. {\bf 98}, 060404 (2007).
\bibitem{AstrakharchikPhaseTransition} G. E. Astrakharchik, J. Boronat, I. L. Kurbakov, and Yu. E. Lozovik, 
Phys. Rev. Lett. {\bf 98}, 060405 (2007).

\bibitem{ODellDipolarHydrodynamics} Duncan H. J. O'Dell, Stefano Giovanazzi, and Claudia Eberlein, 
Phys. Rev. Lett. {\bf 92}, 250401 (2004).
\bibitem{EberleinThomasFermi} Claudia Eberlein, Stefano Giovanazzi, and Duncan H. J. O'Dell, 
Phys. Rev. A {\bf 71}, 033618 (2005).

\bibitem{ODellDepletion} I. E. Mazets, D. H. J. O'Dell, G. Kurizki, N. Davidson, and W. P. Schleich, 
J. Phys. B: At. Mol. Opt. Phys. {\bf 37}, S155 (2004).

\bibitem{NathPhononInstabilityTwoDimDipoles} R. Nath, P. Pedri, and L. Santos, Phys. Rev. Lett. {\bf 102}, 
050401 (2009).

% \bibitem{Rosensweig} M. D. Cowley and R. E. Rosensweig, J. Fluid Mech. {\bf 30}, 671 (1967).
% 
% \bibitem{SurfaceUndulationFerrofluid} Holger Knieling, Reinhard Richter, Ingo Rehberg, Gunar Matthies, and Adrian Lange,
% Phys. Rev. E {\bf 76}, 066301 (2007).

\bibitem{ChanSuperSolid} E. Kim and M. H. W. Chan, Nature {\bf 427}, 225 (2004).

\bibitem{DalibardBKT} Zoran Hadzibabic, Peter Kr\"uger, Marc Cheneau, Baptiste Battelier, and Jean Dalibard, 
Nature {\bf 441}, 1118 (2006).

\bibitem{ExcitonBEC} A. V. Larionov, V. B. Timofeev, P. A. Ni, S. V. Dubonos, I. Hvam, and K. Soerensen, JETP Letters {\bf 75}, 
570 (2003).

\bibitem{MonteCarloExcitons} Yu. E. Lozovik, S. Yu. Volkov, and M. Willander, JETP Letters {\bf 79}, 473 (2004).


\end{thebibliography}
\end{document}